\documentclass[onecolumn,aps,floatfix,final,prl,showpacs,superscriptaddress]{revtex4}
\usepackage[latin9]{inputenc}
\usepackage{amsmath}
\usepackage{amssymb}
\usepackage{graphicx}

\makeatletter

\providecommand{\tabularnewline}{\\}

\@ifundefined{textcolor}{}
{%
 \definecolor{BLACK}{gray}{0}
 \definecolor{WHITE}{gray}{1}
 \definecolor{RED}{rgb}{1,0,0}
 \definecolor{GREEN}{rgb}{0,1,0}
 \definecolor{BLUE}{rgb}{0,0,1}
 \definecolor{CYAN}{cmyk}{1,0,0,0}
 \definecolor{MAGENTA}{cmyk}{0,1,0,0}
 \definecolor{YELLOW}{cmyk}{0,0,1,0}
}

\usepackage{epsfig}

\makeatother

\begin{document}

\title{Possible unconventional superconductivity in substituted BaFe$_{2}$As$_{2}$ 
revealed by magnetic pair-breaking studies}

\author{P. F. S. Rosa \footnote{Corresponding author: pfsrosa@uci.edu}}

\affiliation{Instituto de F\'{i}sica \lq\lq Gleb Wataghin\rq\rq, UNICAMP,
Campinas-SP, 13083-859, Brazil.}

\affiliation{University of California, Irvine, California 92697-4574, USA.}

\author{C. Adriano}

\affiliation{Instituto de F\'{i}sica \lq\lq Gleb Wataghin\rq\rq, UNICAMP,
Campinas-SP, 13083-859, Brazil.}

\author{T. M. Garitezi}

\affiliation{Instituto de F\'{i}sica \lq\lq Gleb Wataghin\rq\rq, UNICAMP,
Campinas-SP, 13083-859, Brazil.}

\author{M. M. Piva}

\affiliation{Instituto de F\'{i}sica \lq\lq Gleb Wataghin\rq\rq, UNICAMP,
Campinas-SP, 13083-859, Brazil.}
\affiliation{Max Planck Institute for Chemical Physics of Solids, N\"{o}thnitzer Str. 40, D-01187 Dresden, Germany.}

\author{K. Mydeen}

\affiliation{Max Planck Institute for Chemical Physics of Solids, N\"{o}thnitzer Str. 40, D-01187 Dresden, Germany.}

\author{T. Grant}

\affiliation{University of California, Irvine, California 92697-4574, USA.}

\author{Z. Fisk}

\affiliation{University of California, Irvine, California 92697-4574, USA.}

\author{M. Nicklas}

\affiliation{Max Planck Institute for Chemical Physics of Solids, N\"{o}thnitzer Str. 40, D-01187 Dresden, Germany.}

\author{R. R. Urbano}

\affiliation{Instituto de F\'{i}sica \lq\lq Gleb Wataghin\rq\rq, UNICAMP,
Campinas-SP, 13083-859, Brazil.}

\author{R. M. Fernandes}

\affiliation{School of Physics and Astronomy, University of Minnesota, Minneapolis,
MN 55455, USA.}

\author{P. G. Pagliuso}

\affiliation{Instituto de F\'{i}sica \lq\lq Gleb Wataghin\rq\rq, UNICAMP,
Campinas-SP, 13083-859, Brazil.}

\date{\today}


\maketitle

\textbf{The possible existence of a sign-changing gap symmetry in BaFe$_{2}$As$_{2}$-derived
superconductors (SC) has been an exciting topic of research in the last few years. To further investigate this subject we combine Electron Spin Resonance (ESR) and pressure-dependent transport
measurements to investigate magnetic pair-breaking effects on BaFe$_{1.9}M_{0.1}$As$_{2}$
($M=$ Mn, Co, Cu, and Ni) single crystals. An ESR signal,
indicative of the presence of localized magnetic moments, is observed
only for $M=$ Cu and Mn compounds, which display very low SC transition temperature ($T_{c}$)
and no SC, respectively. From  the  ESR analysis assuming the absence of bottleneck effects, the microscopic parameters are extracted to show that this reduction of $T_{c}$ cannot be
accounted by the  Abrikosov-Gorkov pair-breaking expression for a sign-preserving
gap function. Our results reveal an unconventional spin- and pressure-dependent pair-breaking effect and impose strong constraints on the pairing symmetry of these materials.}

The Fe-based superconductors (SC) $R$FeAsO ($R=$ La-Gd) and $A$Fe$_{2}$As$_{2}$
($A=$ Ba,$\,$Sr,$\,$Ca,$\,$Eu) have been a topic of intense scientific
investigation since their discovery \cite{Kamihara,Rotter}. In particular,
the semi-metal member BaFe$_{2}$As$_{2}$ (Ba122) displays a spin-density
wave (SDW) phase transition at $139$ K which can be suppressed by hydrostatic 
pressure and/or chemical substitution
(e.g. K, Co, Ni, Cu, and Ru)  inducing a SC phase \cite{Review,Alireza, Review2,Review3,Review4}. Although the proximity to
a SDW state suggests a magnetic-mediated pairing mechanism \cite{Mazin_review,Chubukov_review},
the precise nature and symmetry of the SC state, as well as the microscopic
mechanism responsible for driving the SDW phase towards a SC state,
remain open questions begging for further investigation. 
Importantly, suppressing the SDW phase -- either via applied pressure or 
chemical substitution -- is not sufficient for SC
to emerge \cite{Mn_pressure, Canfield09}. Furthermore, when SC is found, the achieved optimal $T_{c}$ differs dramatically depending on the particular chemical substitution. This difference may be related to the pair-breaking
effect associated with substitutions, which create
local impurity scatterers, particularly when introduced in the FeAs
planes \cite{Paglione}. 

A complete understanding of the impurity pair-breaking
(IPB) effect in the Fe-pnictides is hindered, however, by their
multi-band character and by the absence of quantitative information
about the impurity potential \cite{Mazin_review}. Indeed, the suppression of $T_{c}$
by impurities has been used as an argument in favor of
both a sign-preserving $s^{++}$ state \cite{s++,Kontani2} and a
sign-changing $s^{+-}$ state in Ba122-derived materials \cite{Bang,s+-,Rafael}.
In these analyses, the impurity potential is usually estimated by
the changes in the residual resistivity. However, the latter is sensitive to
the transport scattering rate, which may differ from the quasi-particle
scattering rate related to the suppression of $T_{c}$. Furthermore,
using optimally-doped (OPD) compositions to study the effects of impurities
on $T_{c}$ may introduce additional complications, since
any kind of perturbation will likely drive the system away from the
vicinity of the SDW phase and suppress SC by diminishing the
strength of the pairing interaction instead of breaking the
Cooper pairs \cite{Kontani2,Ni}.

In this paper, we circumvent these issues by combining
macro and microscopic experiments, namely pressure-dependent
transport measurements and electron spin resonance (ESR) in order to investigate
the magnetic IPB effects in BaFe$_{1.9}M_{0.1}$As$_{2}$ ($M=$ Mn,
Co, Cu, and Ni) single crystals slightly below the OPD concentration. A sizeable ESR signal for  $M=$ Mn, Cu samples provides
not only direct evidence for their role as local magnetic
impurities, but it also allows us to extract the averaged exchange
coupling $\langle J^{2}(\mathbf{q})\rangle$ between them and the
Fe $3d$ conduction electrons. The estimated suppression of $T_{c}$ derived
from this quantity, which plays the role of the magnetic impurity potential
in the Abrikosov-Gor'kov (AG) formalism \cite{AG,Skalski}, is found to
be significantly smaller than the observed one, in sharp contrast
to the excellent agreement found previously in borocarbides \cite{Pagliuso4,LaSn,LaAl2}
-- multi-band compounds that display conventional sign-preserving
SC states. Furthermore, we find that pressure strongly enhances
$T_{c}$ of the $M=$ Cu sample, presumably by promoting stronger
Cu--Fe hybridization and consequently suppressing the IPB
effect. Our findings impose strong constraints on the
mechanism responsible for SC  and provide a strong evidence for an
unconventional gap symmetry in these materials.

Fig. 1 displays the in-plane electrical resistivity, $\rho_{ab}$ ($T$), at ambient pressure for the selected
 single crystals. A linear metallic behavior is observed at high-$T$ and 
the SDW phase transition of the parent compound is suppressed for all substitutions. 
 A slight upturn is still present (arrows in Fig. 1), as typically found for substituted
 samples of Ba122 slightly below the OPD concentration \cite{Review}. As $T$ is further decreased, SC emerges with the onset of $T_{c}$, defined as the temperature 
at which $d\rho_{ab}/dT = 0$,
 at $26.1$ K, $22.2$ K, and $3.8$ K for Co, Ni, and Cu substitutions, respectively. 
On the other hand, no $T_{c}$ is observed for $M=$ Mn.

\begin{figure}[!ht]
\includegraphics[width=0.42\textwidth]{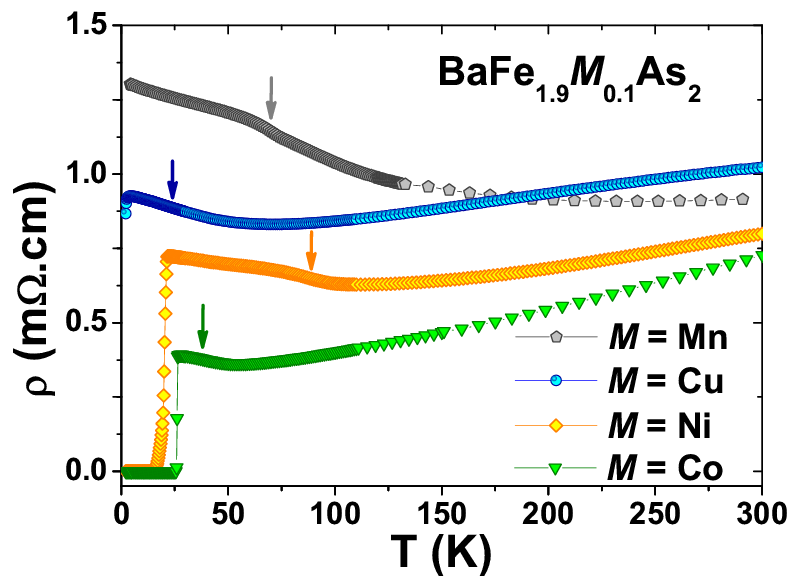} 
\caption{In-plane electrical resistivity, $\rho_{ab}$ ($T$), for BaFe$_{1.9}M_{0.1}$As$_{2}$
($M$ $=$ Mn, Cu, Ni, Co) single crystals. The arrows show the minima of the first derivative in the vicinity of the SDW transition.}
\label{Fig1} 
\end{figure}

Figs. 2a-b show $\rho_{ab}$ ($T$) as a function of pressure for Co
and Ni-substituted compounds. A small increase of $T_{c}$ is observed,
as expected for nearly OPD samples \cite{Review}. For instance,
$T_{c}$ reaches $28.6$ K at $18$ kbar for $M=$ Co, whereas the
 self-flux OPD compound reaches a maximum $T_{c}$  of $\sim 23$ K in the same pressure range,
suggesting that the In-flux samples are of high quality.
On the other
hand, for $M=$ Ni, $T_{c}$ only reaches $24.7$ K. One can speculate that the reason
 the Ni-OPD sample does not achieve $T_{c}\sim29$ K is that it
introduces more disorder than cobalt \cite{Sawatzky, Ideta, Ber}. Indeed, the residual resistivity
is higher for $M=$ Ni. Furthermore, the highest $T_{c}$ found in FeAs-based
SC is obtained through out-of-plane substitution \cite{K,Gd}.

\begin{figure}[!ht]
\hspace*{-0.5cm}
\includegraphics[width=0.52\textwidth]{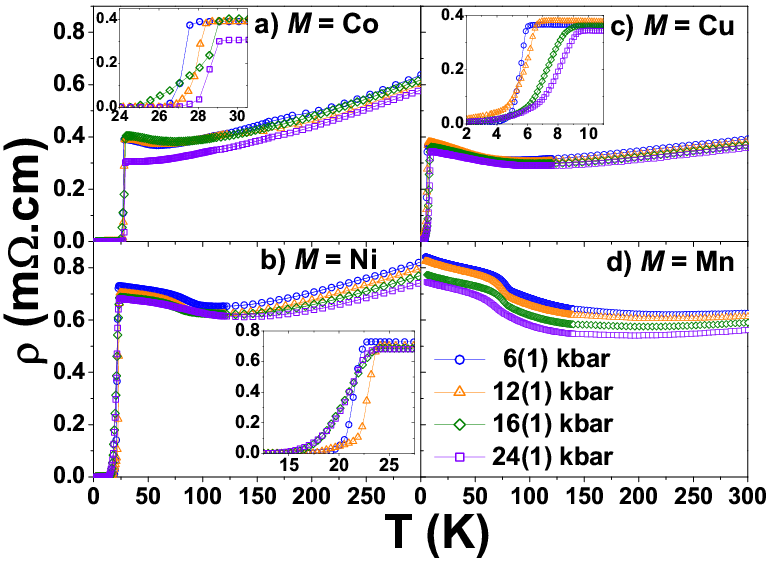} \caption{$\rho_{ab}$($T$) vs. $T$ for BaFe$_{1.9}M_{0.1}$As$_{2}$ ($M=$
Co, Cu, Ni, and Mn) single crystals for $P=5$ \textendash{}$25$ kbar. The insets show the evolution of $T_{c}$ with pressure.}
\end{figure}

Now we turn our attention to the striking behavior of Mn- and
Cu-substituted compounds, shown in Figs. 2c-d. First, we observe
a substantial unexpected enhancement of $T_{c}$ by a factor of $\sim 2.5$ ($T_{c}=10$
K at $24$ kbar) for the Cu-substituted compound. Although an increase of $T_{c}$ is expected for underdoped samples \cite{Ahilan,Drotziger,Yamaichi}, the maximum $T_{c}$ achieved is also expected to be roughly the same as in the OPD sample at ambient pressure. Surprisingly, this is not the case for the studied Cu-substituted compound, which presents $T_{c} = 4.2$ K for the OPD crystal. On the other hand, SC does not emerge for  $M=$ Mn 
up to $P=25$ kbar, in agreement with previous reports \cite{Mn_pressure}.
In addition, there is a drastic
decrease of $\rho_{ab}$($T$) with pressure by a factor of  $\sim3$ for $M=$ Cu and of $\sim1.5$ for $M=$ Mn over all $T$ range (see Fig. 1 for a comparison), suggesting a possible decrease
of the impurity scattering potential.
These results seem to be consistent with a magnetic IPB mechanism
since -- unlike their Co and Ni counterparts -- Mn and Cu substitutions
are expected to introduce local moments. In many compounds, pressure is well known 
 to enhance the
hybridization between the local moments and the conduction electrons \cite{CeCu6,Ce343,CeAl3,Sheila,Hering}.
Such enhancement would suppress the magnetic IPB effect and, consequently, increase
$T_{c}$. As Mn$^{2+}$ has a much higher spin ($S=5/2$) than
Cu$^{2+}$ ($S=1/2$), it is not surprising that the magnetic IPB
is larger for $M=$ Mn, which in turn does not display SC.

To investigate such magnetic IPB scenario, we performed ESR -- a powerful
spin probe technique sensitive to the presence of local moments and
 their coupling to the conduction electrons \cite{Dyson}. In agreement with the expectation that Cu and Mn ions have local moments, our ESR data reveal an intense resonance line
for $M=$ Cu and Mn, but not for $M=$ Co and Ni. Fig. 3 shows the X-Band ESR lines normalized by the concentration of paramagnetic ions at
$T=150$ K for fine powders of gently crushed single crystals. The Lorentzian fitting
of the spectra  reveals a linewidth of $\Delta H=600(60)$ G and a
$g$-value of $g=2.08(3)$ for $M=$ Cu. For $M=$ Mn, $g=2.04(3)$
and the linewidth is slightly larger, $\Delta H=750(80)$ G, indicating
stronger Mn-Mn interactions. Finally, for $M=$ Mn and Co, $g=2.05(3)$
and $\Delta H=670(70)$ G. For all samples, the calibrated number
of resonating spins at room-$T$ is in good agreement with the concentrations obtained
from Energy Dispersive Spectroscopy (EDS). As expected, the ESR intensity, which is proportional to $S(S+1)$, was found to
be roughly twelve times larger for $M=$ Mn samples, as compared to the $M=$ Cu sample.
These results also indicate that the
 oxidation states of Cu and Mn are indeed Cu$^{2+}$ ($S = 1/2$) and Mn$^{2+}$ ($S = 5/2$). In the former case, Cu$^{+}$ ($3d^{10}$ state) would not display an ESR resonance line since it is not a paramagnetic ion. In the case of copper, Cu$^{+}$ ($3d^{10}$ state) would not display an ESR resonance line since it is not a paramagnetic probe with unpaired electrons. In the case of manganese, for Mn$^{3+}$ ($S=2$) and Mn$^{4+}$ ($S=3/2$) ions, one would expect a distinct ESR response (i.e., different g-value and calibrated signal intensity). Consequently, 
one can infer that there is no effective charge doping into the system, as suggested previously both experimentally and theoretically \cite{Bittar, Sawatzky}.
 Furthermore, our ESR results agree with other
indirect probes that also suggest localized Cu$^{2+}$ and Mn$^{2+}$
moments in chemically-substituted Ba122 \cite{Mn_NMR2,Mn_NMR,Mn_ARPES,Kim,Frankovsky}.
We note that the detailed analysis of the ESR data confronted with Eu-substituted
BaFe$_{2-x}M_{x}$As$_{2}$ ($M=$ Co, Ni, Cu, Mn, and Ru) requires further technical discussion. Therefore, it will be the focus of a separated report \cite{MMM,SS}.

\begin{figure}[!ht]
\includegraphics[width=0.45\textwidth,keepaspectratio]{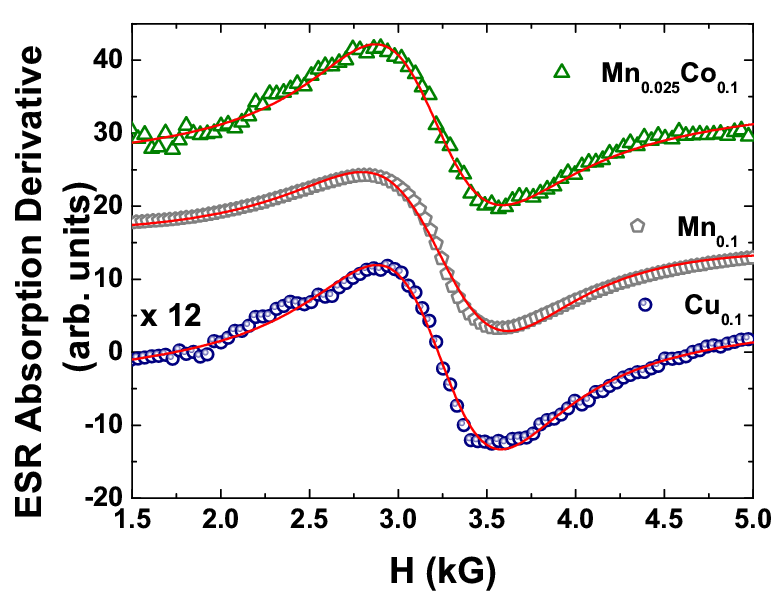} \caption{X-Band ESR lines at $T=150$ K for powdered crystals of BaFe$_{1.9}M_{0.1}$As$_{2}$
($M=$ Cu, Mn). The solid lines are Lorentzian fits to the spectra (sample grain size smaller than the skin depth \cite{Dyson}). It
is worth mentioning that, in order to obtain the ESR signal, the sample
surface must be completely clean and free of In-flux. The ESR signals for
both samples were calibrated at $300$ K using a strong pitch standard sample
with $4.55\times10^{15}$ spins/cm.}
\end{figure}

Besides revealing the presence of localized moments, ESR also allows
us to extract the averaged squared exchange coupling $\langle J^{2}(\mathbf{q})\rangle$
between the localized moments and the conduction electrons from the linear increase
of the linewidth with temperature (Korringa behavior) (see Table I)
\cite{PFSRosa,SS,MMM, Pagliuso4,LaSn,LaAl2}. In a general approach for single-band metals, the thermal broadening $b$ of the ESR linewidth $\Delta H \simeq 1/T_{1}$ is the linear well-known Korringa relaxation defined as $b \equiv \frac{d\left(\Delta H\right)}{dT} = \frac{\pi k_B}{g \mu_B} \langle J^2_{fs}({\textbf{q}}
)\rangle\eta^2\left(E_F\right)~\frac {K(\alpha)}{(1 - \alpha)^2}$, \noindent \cite{ref19}. Here, $\langle J_{fs}(\textbf{q})^{2}\rangle^{1/2}$ is the effective exchange interaction between the local moment and the conduction electrons ($ce$) in the presence of $ce$ momentum
transfer averaged over the whole Fermi surface (FS) \cite{ref21}, $\eta(E_F)$ is the ``$bare$" density of states (DOS)
for one spin direction at the Fermi level, $g$ is the local moment $g$-value and $K(\alpha)$ is the Korringa exchange enhancement factor due to electron-electron exchange interaction \cite{ref28,ref29}. In the present analysis, we found empirically that `$bottleneck$" and ``$dynamic$" effects are not present \cite{ref20}. When ``$dynamic$" effects are present the g-values are usually strongly $T$-dependent, which is not observed in our experimental data. Moreover, when `$bottleneck$" effects are relevant the Korringa rate $b$ decreases with increasing concentration of the magnetic ions. However, in our data, we observe that spin-spin interaction dominates the entire temperature range for dilute concentrations of Mn and Cu ions. In addition, bottleneck effects are not observed in Eu-substituted BaFe$_{2}$As$_{2}$ \cite{PFSRosa}, indicating that FeAs-based compounds are intrinsically unbottlenecked systems likely due to fast relaxation rates between the 3d conduction electrons and the lattice. In fact, recent ultrafast spectroscopy measurements have found a very large spin-lattice coupling in BaFe$_{2}$As$_{2}$ \cite{Patz}. Finally, even if bottleneck effects were present, they alone would hardly be able to account for the enormous difference between $J_{ESR}$ and $J_{AG}$ observed here.

The key point here is that this parameter $\langle J^{2}(\mathbf{q})\rangle$
is the same one determining the suppression of $T_{c}$ by magnetic
impurities within the AG formalism \cite{AG,Skalski}. To estimate whether the extracted value of $\langle J^{2}(\mathbf{q})\rangle_{ESR}$
for $M=$ Cu and Mn compounds can account for the observed suppression
 of $T_{c}$ in this formalism, we consider the ``conventional case'', where
the gap function has the same amplitude and sign across the entire
Brillouin zone. This is the scenario in which magnetic impurities
have the strongest effect on $T_{c}$ -- in fact, introducing anisotropies
 in the gap function would make the magnetic pair-breaking
effect weaker \cite{Golubov_Mazin,Openev}.  In this situation, we
have \cite{Skalski}:

\begin{equation}
\mathrm{ln}\left(\frac{T_{c,0}}{T_{c}}\right)=\psi\left(\frac{1}{2}+\frac{1}{2\pi T_{c}\tau_{s}}\right)-\psi\left(\frac{1}{2}\right),\label{digamma}
\end{equation}

\noindent where $\psi(x)$ is the digamma function, $T_{c,0}$ is
the transition temperature in the absence of magnetic impurities,
and $\tau_{s}^{-1}=\frac{\pi}{2}\Delta c\, \eta(E_{F})\langle J^{2}(\mathbf{q})\rangle S\left(S+1\right)$
 is the magnetic scattering rate. Here, $\eta(E_{F})$ is the density of
states per spin at the Fermi level, $\Delta c$ is the magnetic impurity
concentration, and $S$, the spin of the localized moment. 
Given that $\Delta c < 0.1$ in our samples, we can perform a series expansion of eq. \ref{digamma}
and obtain the simplified expression:

\begin{equation}
\left|\frac{\Delta T_{c}}{\Delta c}\right|=\frac{\pi^{2}}{8}\eta (E_{F})\langle J^{2}(\mathbf{q})\rangle S(S+1),
\end{equation}

\noindent with $\Delta T_{c}^{\mathrm{}}=T_{c}-T_{c,0}$. The value for $\eta(E_{F})$ is extracted from the linear coefficient
of the low-temperature specific heat $\gamma$, yielding $\eta(E_{F})=3.34$
states/eV.spin.FU for one mole, which is the same for all compounds
\cite{Rotter,SS}. Small variations of $\gamma$ across different
compositions would not alter our main conclusions, as we discuss below. Moreover, the nearly constant ESR $g$-shift value found for the various ESR probes as a function of different chemical substitutions in Ba122 compounds is a strong evidence that the density of states at the Fermi level is nearly the same for pure Ba122 and for all studied compounds \cite{SS}.
The choice of $T_{c,0} = 26$ K is a more subtle issue. Since the dependence
of the magnetic SDW transition temperature with $x$ in the phase diagrams
of the BaFe$_{2-x}M_{x}$As$_{2}$ compounds is nearly identical for $M=$
Co, Ni, Cu \cite{Ni}, if one assumes that superconductivity is governed
by fluctuations associated with the normal state, then one would expect
that the optimal $T_{c}$ values of these three samples would be very
similar. Indeed, this is the case for $M=$ Co and Ni, which also
display similar maximum values of $T_{c}$ under pressure. However,
for $M=$ Cu the value of $T_{c}$ is significantly smaller -- but
this sample displays an ESR signal, unlike $M=$ Co and Ni. We therefore
assume that $T_{c,0}$ of the $M=$ Cu sample, and also of the $M=$
Mn sample, is approximately the same as the $T_{c}$ value of the
optimally doped $M=$ Co and Ni samples, where magnetic pair-breaking
is absent, according to our ESR analysis. 
Moreover, theoretical and experimental reports have shown
that there is no effective doping in this class of materials \cite{Bittar, Sawatzky} and that the suppression of $T_{SDW}$ is given by structural parameters \cite{KH, Granado}. As such, in the
absence of pair-breaking effects, a given structural change would lead to the same suppression of the SDW state (and the consequent emergence of SC), independent on the
particular transition metal substitution. We will return to this assumption
below.

With these assumptions, we can thus estimate the magnetic pair-breaking impurity
potential $\langle J^{2}(\mathbf{q})\rangle_{\mathrm{AG}}$ that would
be necessary to cause the observed suppression of $T_{c}$ for three
different samples, namely, BaFe$_{1.9}$Cu$_{0.1}$As$_{2}$, BaFe$_{1.88}$Mn$_{0.12}$As$_{2}$,
and the mixed doping BaFe$_{1.895}$Co$_{0.100}$Mn$_{0.005}$As$_{2}$ compounds.
The results are shown in Table I, and reveal a remarkable
disagreement (of two orders of magnitude) between $\langle J^{2}(\mathbf{q})\rangle_{\mathrm{AG}}$
and the experimentally measured $\langle J^{2}(\mathbf{q})\rangle_{\mathrm{ESR}}$.
This is in sharp contrast to the borocarbide multi-band compounds
Lu$_{1-x}$Gd$_{x}$Ni$_{2}$B$_{2}$C and Y$_{1-x}$Gd$_{x}$Ni$_{2}$B$_{2}$C,
as well as to La$_{1-x}$Gd$_{x}$Sn$_{3}$, all of which display
conventional pairing symmetry. For these materials, as discussed in
Refs. \cite{Pagliuso4,LaSn}, the calculated $\langle J^{2}(\mathbf{q})\rangle_{\mathrm{AG}}$
and the $\langle J^{2}(\mathbf{q})\rangle_{\mathrm{ESR}}$ extracted from ESR experiments 
are in very good agreement, as expected for a conventional SC. We note that in these compounds, because of the presence
of rare earth elements, one needs to properly rewrite the Abrikosov-Gor'kov
Equation \ref{digamma} by replacing $S\left(S+1\right)$ for $\left(g_{J}-1\right)^{2}J\left(J+1\right)$,
where $J=S+L$ is the total spin.

The huge difference between $\langle J^{2}(\mathbf{q})\rangle_{\mathrm{AG}}$
and $\langle J^{2}(\mathbf{q})\rangle_{\mathrm{ESR}}$ is clearly
robust against small variations of $T_{c,0}$ and $\eta(E_{F})$.
As explained above, these conclusions rely on the assumptions that
(i) the Abrikosov-Gor'kov formalism is valid and (ii) similar normal-state
phase diagrams should give similar superconducting transition temperatures.
To shed light on these possible issues, we also present in Table I
the results for the mixed doping BaFe$_{1.895}$Co$_{0.100}$Mn$_{0.005}$As$_{2}$
compound \cite{MMM}. The very small Mn concentration
makes the AG formalism more reliable, and the fact that the compound
without Mn substitution displays a superconducting transition temperature
of $26$ K directly determines $T_{c,0}=26$ K. As shown in the Table,
both $\langle J^{2}(\mathbf{q})\rangle_{\mathrm{AG}}$ and $\langle J^{2}(\mathbf{q})\rangle_{\mathrm{ESR}}$
values are very close to those of the BaFe$_{1.9}$Cu$_{0.1}$As$_{2}$
and BaFe$_{1.88}$Mn$_{0.12}$As$_{2}$ samples, displaying a deviation
of two orders of magnitude.

\begin{table*}[!htb]
\label{Tab_AG} \caption{Experimental and calculated parameters for BaFe$_{1-x}M_{y}$As$_{2}$
(this work) and conventional SC (refs. \cite{Pagliuso4,LaSn})}
\begin{centering}
\begin{tabular}{|c||c|c|c|c|c|c|}
\hline 
Sample  & c (\%)  & g$_{\mathrm{ESR}}$  & $\mid\Delta T_{c}^{\mathrm{exp}}\mid$ (K)  & $T_{c,0}$ (K)  & $\langle J^{2}(\mathbf{q})\rangle_{\mathrm{ESR}}^{1/2}$ (meV)  & $\langle J^{2}(\mathbf{q})\rangle_{\mathrm{AG}}^{1/2}$ (meV) \tabularnewline
\hline 
BaFe$_{1.9}$Cu$_{0.1}$As$_{2}$  & 5  & 2.08(3)  & 22  & 26  & 1.2(5)  & 111(10) \tabularnewline
BaFe$_{1.88}$Mn$_{0.12}$As$_{2}$  & 6  & 2.05(2)  & $\geq$ 26  & 26  & 0.7(5)  & $\geq$ 32(3) \tabularnewline
BaFe$_{1.895}$Co$_{0.100}$Mn$_{0.005}$As$_{2}$  & 0.25  & 2.06(2)  & 10  & 26  & 0.8(5)  & 98(9) \tabularnewline
Lu$_{1-x}$Gd$_{x}$Ni$_{2}$B$_{2}$C  & 0.5  & 2.035(7)  & $\approx$0.3  & 15.9  & 10(4)  & 11(1) \tabularnewline
Y$_{1-x}$Gd$_{x}$Ni$_{2}$B$_{2}$C  & 2.1  & 2.03(3)  & $\approx$0.9  & 14.6  & 9(3)  & 10(1) \tabularnewline
La$_{1-x}$Gd$_{x}$Sn$_{3}$  & 0.4  & 2.010(10)  & $\approx$0.5  & 6.4  & 20(2)  & $\approx$20(2) \tabularnewline
\hline 
\end{tabular}
\par\end{centering}
\end{table*}

Our findings have important consequences for the understanding of
the superconductivity in the Fe-pnictides. The fact that $\langle J^{2}(\mathbf{q})\rangle_{\mathrm{ESR}}\ll\langle J^{2}(\mathbf{q})\rangle_{\mathrm{AG}}$
implies that the Abrikosov-Gor'kov magnetic IPB alone cannot account for the suppression
of $T_{c}$. The latter must therefore be related to an unconventional magnetic IPB which must be strongly associated with the local Cu$^{2+}$ and Mn$^{2+}$ spins. In addition, these 
substitutions could also present a stronger
nonmagnetic IPB effect responsible for part of the
observed suppression of $T_{c}$. This also favors a non-conventional
sign-changing gap function over the more conventional sign-preserving
one, since in the latter case the effects of nonmagnetic
IPB are expected to be weak. We note that in a $s_{+- }$ superconductor, non-magnetic pair-breaking can be weak dependent on the ratio between intra and inter-band scattering \cite{s+-,Efremov}. Furthermore, it is also possible that the substitution
of $M=$ Cu, Mn affects directly the pairing interaction, besides promoting
pair-breaking. Interestingly, for $M=$ Mn substitution, along with the
usual SDW-type fluctuations, N\'{e}el-type fluctuations are also observed
by inelastic neutron scattering \cite{Mn_neutron}. Even when these
N\'{e}el fluctuations are weak and short-ranged, they have been shown
theoretically to strongly suppress $T_{c}$ \cite{Rafael2}. We note that our results are
in agreement with recent measurements on LiFeAs employing angle resolved photoemission spectroscopy (ARPES) combined with quasiparticle interference (QPI) by means of scanning tunneling microscopy/spectroscopy (STM/STS) \cite{STM}.

Finally, we comment on the effects of pressure on $T_{c}$, summarized
in Fig. 4. For the Co and Ni substitutions, the rate d$T_{c}$/d$P$ is $\sim0.1$
K/kbar and the application of pressure has little effect on $T_{c}$.
Strikingly, this rate is three times larger for the $M=$ Cu sample,
while for $M=$ Mn, no SC is observed. We argue that these results are linked to the magnetic pair-breaking
discussed above. In particular, because pressure increases the hybridization
between the Cu $3d$ bands and conduction electron bands, the copper
bands become more itinerant, progressively losing their local moment
character and consequently suppressing the magnetic IPB effect. Therefore,
it is not surprising that the pure BaCu$_{2}$As$_{2}$ is a Pauli
paramagnet with completely delocalized Cu $3d$ bands and no phase
transition. Within this scenario, the fact that the Mn compounds do
not display SC would follow from the fact that Mn$^{2+}$ has a spin
value five times larger than Cu$^{2+}$. Interestingly, if the magnetic
IPB mechanism is suppressed by pressure, $T_{c}$ is, in principle,
unconstrained to increase up to a maximum defined by the local distortions
that the $M$-substitution creates. For Cu-substituted samples, it remains to be confirmed  whether applying higher pressures with Diamond Anvil Pressure cells would further enhance or even suppress $T_{c}$ in the impurity pair-breaking regime.

To make this reasoning more quantitative, we assume that the enhancement
of $T_{c}$ caused by the magnetic IPB suppression with pressure follows a phenomenological
expression of the form $\Delta T_{c}=S(S+1)(a-bP)$, where $a$ and
$b$ are free parameters and $P$, applied pressure. The linear dependence
with pressure is motivated by the same typical dependence of the Kondo
temperature ($T_{K}$) on pressure in several Ce-based heavy fermion
compounds \cite{CeCu6,Ce343,CeAl3,Sheila,Hering}. This linear regime
can be applied to the $M=$ Cu ($S=1/2$) compound 
in the IPB region slightly below the optimally-doped concentration, where the spin fluctuactions 
are nearly constant as a function of pressure. This procedure allows one to obtain the linear fit 
to the experimental data (solid line) displayed
in Fig. 4. 
On the other hand, for $M=$ Cu compounds in the optimally-doped or overdoped regions, 
the spin fluctuation suppression starts to play an important role
and would overcome the latter linear increase of $T_{c}$. A detailed study on the effects of Cu substitution in critical current measurements is presented in Ref. \cite{Garitezi2}.
Now, by constraining the same linear dependence for $M=$ Mn and changing only the spin value to $S=5/2$, we obtain a lower limit
for the critical pressure $P_{c}\sim 66$ kbar necessary for the emergence
of SC (dashed line in Fig. 4). This $P_{c}$ value is in good agreement with the experimental absence of SC
in the $M=$ Mn compounds up to
$25$ kbar (see Fig. 2d), also in agreement with previous reports \cite{Mn_pressure}.

\begin{figure}[!ht]
\hspace*{-0.4cm} \includegraphics[width=0.52\textwidth,keepaspectratio]{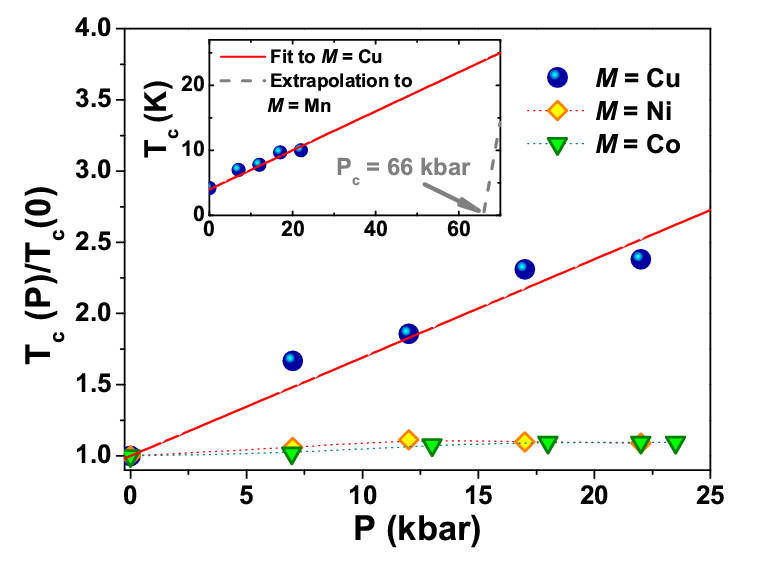}
\caption{Phase diagram for BaFe$_{2-x}M_{x}$As$_{2}$ ($M =$ Co, Cu, and Ni) single crystals as a function of pressure.
The dotted lines are guide to the eyes for the SC domes. The
 linear fit for the $M=$ Cu compound (solid line) was obtained from the phenomenological 
expression $\Delta T_{c}=S(S+1)(a-bP)$. 
Using the same expression and $S=5/2$, we obtain the dashed line for the $M=$ Mn compound.}
\label{PDP} 
\end{figure}

In conclusion, we have demonstrated the contrasting behavior of hydrostatic
pressure effects on nearly OPD BaFe$_{2-x}M_{x}$As$_{2}$ ($M=$
Co, Cu, and Ni) high-quality single crystals grown from In-flux method.
The striking enhancement of $T_{c}$ with pressure for $M=$ Cu and the
existence of a Cu$^{2+}$ ESR line provide strong evidence of a spin-dependent
pair-breaking mechanism strongly suppressed by pressure, suggesting 
an increase of hybridization between the Cu $3d$ bands
and the conduction electron bands. More interestingly, by using the
magnetic impurity potential extracted from the ESR analysis in the absence of \textit{bottleneck} effects, we find
that the Abrikosov-Gor'kov pair-breaking mechanism, applied to a conventional
sign-preserving pairing state, cannot account for the observed suppression
of $T_{c}$ in the Cu and Mn-substituted compounds. This result not
only implies that the suppression of $T_{c}$ in these samples is
due to other mechanisms, but also that an unconventional pairing state
is more likely to be realized. 

\subsection*{Methods}

Single crystals of BaFe$_{1.9}M_{0.1}$As$_{2}$ ($M=$ Mn, Co, Cu,
and Ni) were grown using In-flux as described elsewhere \cite{Garitezi}.
The crystals were checked by x-ray powder diffraction and submitted
to elemental analysis using a commercial EDS microprobe. In-plane
electrical resistivity measurements were performed using a standard four-probe
method and a self-contained piston-cylinder type Be-Cu pressure cell,
with a core of hardened NiCrAl alloy. ESR spectra were taken in
a commercial ELEXSYS 500 X-band ($\nu=9.5$ GHz) spectrometer equipped
with a continuous He gas-flow cryostat.

\subsection*{Author contributions}

P.F.S.R., C.A., and T.M.G. have grown the single crystals and performed pressure dependent transport measurements. P.F.S.R. performed ESR measurements. T. G. performed EDS measurements. R. M. F. performed theoretical analyses. P.F.S.R., C.A., T.M.G, T.G., Z.F., R.R.U., R.M.F., P.G.P. discussed the data and reviewed the manuscript.

\subsection*{Additional information}

Competing financial interests: The authors declare no competing financial interests.

\subsection*{\label{sec:acknowlegment} Acknowledgment}

This work was supported by FAPESP-SP, AFOSR MURI, CNPq and FINEP-Brazil. RMF acknowledges the financial support of the APS-SBF Brazil-US Professorship/Lectureship Program.

%
%
%
%
%
Using the same expression and $S=5/2$, we obtain the dashed line for the $M=$ Mn compound.\\

\end{document}